\documentclass[12pt]{article}
\begin{document}
\title{\normalsize \hfill UWThPh-2001-21 \\[1cm] \LARGE
Softly broken lepton numbers \\
and maximal neutrino mixing}
\author{Walter Grimus\thanks{E-mail: grimus@doppler.thp.univie.ac.at} \\
\setcounter{footnote}{3}
\small Institut f\"ur Theoretische Physik, Universit\"at Wien \\
\small Boltzmanngasse 5, A--1090 Wien, Austria \\*[3.6mm]
Lu\'{\i}s Lavoura\thanks{E-mail: balio@cfif.ist.utl.pt} \\
\small Universidade T\'ecnica de Lisboa \\
\small Centro de F\'\i sica das Interac\c c\~oes Fundamentais \\
\small Instituto Superior T\'ecnico, P--1049-001 Lisboa, Portugal \\*[4.6mm] }

\date{20 July 2001}

\maketitle

\begin{abstract}
We consider lepton mixing
in an extension of the Standard Model
with three right-handed neutrino singlets.
We require that the three lepton numbers $L_e$, $L_\mu$, and $L_\tau$
be separately conserved in the Yukawa couplings,
and we assume that they are {\em softly\/} broken
only by the Majorana mass matrix $M_R$
of the neutrino singlets.
In this framework,
where lepton-number breaking occurs at a scale
much higher than the electroweak scale,
deviations from family-lepton-number conservation are
calculable and finite,
and lepton mixing stems exclusively from $M_R$.
We then show that a discrete symmetry exists such that,
in the lepton mixing matrix $U$,
maximal atmospheric neutrino mixing together with $U_{e3}=0$
can be obtained {\em naturally}.
Alternatively,
if one assumes that there are two different scales in $M_R$
and that the lepton number $\bar L = L_e - L_\mu - L_\tau$ is
conserved in between them,
then maximal solar neutrino mixing
follows naturally.
If both the discrete symmetry
and intermediate $\bar L$ conservation are introduced,
bimaximal mixing is achieved.
\end{abstract}

\newpage

\section{Introduction} \label{introduction}

At present the positive results of some experimental searches
for neutrino oscillations
provide the only glimpse of physics beyond the Standard Model (SM).
In particular,
the results of the atmospheric and solar neutrino experiments
have a natural explanation in terms of neutrino oscillations
\cite{pontecorvo} and,
therefore,
in terms of non-zero light-neutrino masses $m_k$ ($k=1,2,3$)
and lepton mixing.
Atmospheric neutrino results \cite{SK}
are well fitted by a mass-squared difference
$\Delta m^2_\mathrm{atm} = |m_3^2-m_2^2|
\simeq 3 \times 10^{-3}$ eV$^2$ and a mixing angle $\psi \sim 45^\circ$.
The situation for solar neutrinos is not that clear-cut:
several viable explanations of the solar-neutrino deficit seem to exist
(for recent analyses see \cite{bahcall,GG,GG-valle,fogli};
for the latest experimental results of Super-Kamiokande see \cite{latest}).
The large-mixing-angle (LMA) MSW \cite{MSW} solution
has $\Delta m^2_\odot = |m_2^2 - m_1^2|
\simeq 3 \times 10^{-5}$ eV$^2$ and a large mixing angle $\theta$;
the small-mixing-angle (SMA) MSW solution features
$\Delta m^2_\odot \simeq 5 \times 10^{-6}$ eV$^2$
and $\tan^2 \theta \simeq 6 \times 10^{-3}$;
finally,
the LOW solution has $\Delta m^2_\odot \simeq 10^{-7}$ eV$^2$
and a large mixing angle,
possibly reaching $45^\circ$.
(All these numbers have been taken from \cite{GG}.)

The problem of neutrino masses and lepton mixing
can be separated into two questions.
The first one concerns mechanisms for achieving neutrino masses
which are much smaller than the charged-lepton masses;
the seesaw mechanism \cite{seesaw}
figures as a prominent solution to this problem
and will be adopted in the present paper.
The second question concerns the specific features
of the lepton mixing matrix $U$;
in particular,
how can an atmospheric mixing angle close to $45^\circ$,
and a very small $U_{e3}$ \cite{chooz} (see also \cite{GG-valle,fogli}),
be explained.
This second question has been viewed
from many different angles (see for instance \cite{barr})
and no preferred solutions have emerged yet.
In particular,
it has been observed \cite{barbieri} that the lepton number
$\bar L = L_e - L_\mu - L_\tau$,
which must be broken at some stage \cite{nir}
to obtain $\Delta m^2_\odot \neq 0$,
leads to maximal {\em solar\/} mixing---see also \cite{lavoura00}
and the references therein.
This is certainly an interesting observation,
but it may be argued that explaining maximal {\em atmospheric\/} mixing
should have a higher priority than explaining maximal solar mixing---since
the latter is not yet firmly established by experiment.
Now,
it has proven difficult to obtain maximal atmospheric mixing
in a {\em natural\/} fashion
(i.e., protected by some symmetry),
in particular because the charged-lepton mass matrix
may be non-diagonal and blur the picture of mixing
following from the neutrino mass matrix.

In the present paper we take up and expand the idea of using lepton numbers
for enforcing desired features upon the mixing matrix $U$.
We start from the lepton sector of the SM
enlarged by three right-handed neutrino singlets
with Majorana mass terms given by the mass matrix $M_R$.
The scalar sector of our theory consists only of SU(2) doublets
and we do not introduce other scalar multiplets like singlets or triplets.
The main idea of our work is to impose separate conservation
of the family lepton numbers $L_e$, $L_\mu$, and $L_\tau$
(for a recent review on the status of lepton numbers see \cite{bilenky})
and to allow them to be broken only {\em softly},
a breaking which may occur exclusively in $M_R$.
Thus,
the family lepton numbers are broken by mass terms of dimension three
at a scale much higher than the electroweak scale.
Under these conditions,
lepton-number-breaking processes have calculable and finite amplitudes.
This framework has the advantage that lepton mixing
originates solely in the neutrino Majorana mass matrix $M_R$;
contributions to $U$ from the charged-lepton mass matrix
are naturally forbidden at tree level
by the assumed conservation of the family lepton numbers
in all terms of dimension four in the Lagrangian.
In this framework,
we then show that it is possible
to impose a discrete symmetry which leads to maximal atmospheric mixing
and to $U_{e3}=0$
(i.e., decoupled solar and atmospheric neutrino oscillations \cite{giunti}).
Alternatively,
it is possible to have two different mass scales $m_R \ll \bar m_R$ in $M_R$
such that for energies in between $m_R$ and $\bar m_R$
the lepton number $\bar L$ is conserved;
we then arrive at a scenario
with maximal solar mixing.\footnote{An earlier variant of this idea,
with two right-handed neutrino singlets only,
was proposed in \cite{lavoura00}.}
Both scenarios may be combined to yield bimaximal neutrino mixing
\cite{bimaximal}.

In section~\ref{framework} we discuss the SM
with family lepton numbers broken exclusively
by the Majorana mass terms of right-handed neutrino singlets.
We show that flavour-changing neutral interactions are not enhanced
in spite of the large right-handed mass scale.
In section~\ref{maxatm} we introduce a model
which naturally yields maximal atmospheric mixing,
whereas in section~\ref{maxsolar}
we discuss intermediate $\bar L$ conservation and maximal solar mixing.
Both features are combined in section~\ref{bimax}.
The conclusions of this paper are found in section~\ref{conclusions}.

\section{The framework} \label{framework}

The framework within which we develop our models
consists of the lepton sector of the SM with three families
together with three right-handed neutrino singlets $\nu_R$.
We allow for an arbitrary number $n_H$ of Higgs doublets $\phi_j$
($j=1, \ldots, n_H$)
and use the notation
\begin{equation}
\phi_j = \left( \begin{array}{c}
\varphi_j^+ \\ \varphi_j^0 \end{array} \right)
\quad {\rm and} \quad
\left\langle 0 \left| \varphi_j^0 \right| 0 \right\rangle
= \frac{v_j}{\sqrt{2}}\, .
\end{equation}
We first compile various well-known formulae
to fix the notation.
The right-handed neutrino singlets have a Majorana mass term
\begin{equation} \label{LM}
\mathcal{L}_M = \frac{1}{2}\, \nu_R^T C^{-1} \! M_R^* \nu_R -
\frac{1}{2}\, \bar \nu_R M_R C \bar \nu_R^T\, ,
\end{equation}
where $M_R$ is symmetric.
We define the left-handed neutrino singlets
$\nu^\prime_L \equiv C \bar \nu_R^T$,
then
\begin{equation} \label{LM2}
\mathcal{L}_M = \frac{1}{2}\, \nu^{\prime T}_L C^{-1} \! M_R \nu^\prime_L -
\frac{1}{2}\, \bar \nu^\prime_L M_R^* C {\bar \nu}_L^{\prime T}.
\end{equation}
The Yukawa Lagrangian of the leptons is given by
\begin{equation} \label{lyukawa}
\mathcal{L}_{\rm Y} = - \sum_{j=1}^{n_H} \left[ \bar \ell_R
\left( \begin{array}{cc} \varphi_j^-, & {\varphi_j^0}^\ast \end{array} \right)
\Gamma_j
+ \bar \nu_R
\left( \begin{array}{cc} \varphi_j^0, & - \varphi_j^+ \end{array} \right)
\Delta_j
\right]
\left( \begin{array}{c} \nu_L \\ \ell_L \end{array} \right)
+ \mbox{h.c.}
\end{equation}
Defining the charged-lepton mass matrix $M_\ell$
and the Dirac neutrino mass matrix $M_D$ as
\begin{equation} \label{Ml}
M_\ell = \frac{1}{\sqrt{2}} \sum_j v_j^\ast \Gamma_j
\quad {\rm and} \quad
M_D = \frac{1}{\sqrt{2}} \sum_j v_j \Delta_j\, ,
\end{equation}
respectively,
we have from $\mathcal{L}_{\rm Y}$ the following mass terms
for the charged
leptons and for the neutrinos:
\begin{equation}
\mathcal{L}_{{\rm Y}_{\rm mass}}
= - \bar \ell_R M_\ell \ell_L
+ \frac{1}{2}\, \nu_L^{\prime T} C^{-1} M_D \nu_L
+ \frac{1}{2}\, \nu_L^T C^{-1} M_D^T \nu^\prime_L
+ \mbox{h.c.}
\end{equation}
The Dirac and Majorana mass terms for the neutrinos may be written together as
\begin{equation}\label{majorana}
\frac{1}{2} \left( \begin{array}{cc} \nu_L^T, & \nu_L^{\prime T}
\end{array} \right) C^{-1} \mathcal{M}_{D+M}
\left( \begin{array}{c} \nu_L \\ \nu_L^\prime \end{array} \right)
+ \mbox{h.c.}
\quad \mbox{with} \quad
\mathcal{M}_{D+M} =
\left( \begin{array}{cc} 0 & M_D^T \\ M_D & M_R
\end{array} \right).
\end{equation}
We assume that there are (at least) two mass scales in the theory:
the electroweak scale,
i.e.,
the one of the vacuum expectation values $v_j$,
and the scale of the Majorana mass terms,
i.e.,
the one of the eigenvalues of $\sqrt{M_R^* M_R}$.
The latter scale is assumed to be much higher than the electroweak scale.
Then,
the neutrino fields participating in the weak interaction,
$\nu_L$,
have effective Majorana mass terms given by
\begin{equation} \label{Lnu}
\mathcal{L}_m = \frac{1}{2}\,
\nu_L^T C^{-1} \mathcal{M}_\nu \nu_L + \mbox{h.c.}\, ,
\end{equation}
with the (approximate) $3 \times 3$ seesaw mass matrix \cite{seesaw}
\begin{equation} \label{seesaw}
\mathcal{M}_\nu = -M_D^T M_R^{-1} M_D\, .
\end{equation}
The $6 \times 6$ Majorana mass matrix of eq.~(\ref{majorana})
is diagonalized by the unitary matrix $\mathcal{U}$ such that
\begin{equation} \label{UU}
\mathcal{U}^T \mathcal{M}_{D+M}\, \mathcal{U} =
\hat m = \mbox{diag} \left( m_1, \ldots , m_6 \right),
\end{equation}
where $m_{1,2,3}$ are seesaw-suppressed whereas $m_{4,5,6}$ are very large.
It is useful to decompose $\mathcal{U}$ into two $3 \times 6$ matrices
$U_L$ and $U_R$ in the following way:
\begin{equation}\label{UU1}
\mathcal{U} =
\left( \begin{array}{c} U_L \\ U_R^* \end{array} \right)
\simeq \left( \begin{array}{cc}
\mathbf{1} & M_D^\dagger {M_R^*}^{-1} \\
-M_R^{-1} M_D & \mathbf{1}
\end{array} \right)
\left( \begin{array}{cc} V & 0 \\ 0 & W^* \end{array} \right),
\end{equation}
where,
in the second part of the equation,
$\mathcal{U}$ has been given to leading order
in the inverse of the high scale \cite{valle}.
At that order,
the $3 \times 3$ unitary matrices $V$ and $W$
are defined by
\begin{equation}\label{VW}
V^T \mathcal{M}_\nu V = \mbox{diag} \left( m_1, m_2, m_3 \right)
\quad \mbox{and} \quad
W^\dagger M_R W^* = \mbox{diag} \left( m_4, m_5, m_6 \right).
\end{equation}

The foundation of our models lies in the assumption
of the conservation of {\em all\/} three lepton numbers $L_\alpha$
($\alpha = e,\mu,\tau$)
in the Yukawa couplings and,
in general,
in all terms of the Lagrangian with dimension four.
This means that the three right-handed charged leptons $\ell_R$
may from the start receive generation labels---we call them $e_R$,
$\mu_R$,
and $\tau_R$.
Similarly,
each of the three right-handed neutrinos $\nu_R$
carries one unit of the corresponding lepton number $L_\alpha$,
and is accordingly named $\nu_{eR}$,
$\nu_{\mu R}$,
or $\nu_{\tau R}$.
Finally,
the three left-handed lepton doublets
$(\nu_L, \hspace*{1mm} \ell_L)^T$
will be called $D_e$,
$D_\mu$,
and $D_\tau$.
The Yukawa coupling matrices $\Gamma_j$ and $\Delta_j$,
and the mass matrices $M_\ell$ and $M_D$,
are all simultaneously diagonal.
The generation lepton numbers $L_\alpha$
are broken explicitly {\em but softly\/} only
by the dimension-three mass terms in $\mathcal{L}_M$ of
eq.~(\ref{LM}).\footnote{The renormalization-group evolution
of the Yukawa-coupling matrices $\Gamma_j$ and $\Delta_j$
does not introduce flavour-changing elements \cite{weinberg}.}

As $M_\ell$ is diagonal,
and since in our model there are three charged leptons
together with six neutrinos,
the $3 \times 6$ matrix $U_L$ in eq.~(\ref{UU1}) is the lepton mixing matrix.
The part of the mixing matrix relevant for the light neutrinos
is approximately unitary and is given by $V$;
up to a phase convention,
this matrix is usually called the neutrino mixing matrix $U$,
i.e.,
\begin{equation}\label{V}
V = e^{i\hat \alpha} U e^{i\hat \beta} \,,
\end{equation}
where the diagonal phase matrix $e^{i \hat \alpha}$ has no physical meaning
(it may be absorbed in the phases of the charged-lepton fields)
while the diagonal phase matrix $e^{i \hat \beta}$ contains phases
which appear only in the lepton-number-violating processes
typical of the Majorana character of the neutrinos.

We claim that in this framework,
in any process,
the deviation from family-lepton-number conservation
proceeds in a finite and calculable way,
controlled by the elements of $M_R$.
In particular,
radiative corrections introduce flavour-changing interactions
of the neutral scalars
due to $\mathcal{L}_M$.
At one-loop level,
the logarithmic term induced by charged-Higgs exchange,
which contains the leading term,\footnote{This term originates
in the vertex correction with the charged scalars in the loop.
The flavour-changing interactions will be studied in detail
in a forthcoming paper \cite{future}.}
is given by
\begin{equation}\label{1loop}
\Delta \Gamma_{j,FC} = \frac{- 1}{16 \sqrt{2}\, \pi^2}\,
\sum_{i=1}^{n_H} \Gamma_i \Delta_j^\dagger
U_R \ln \left( \hat m^2 / m^2 \right) U_R^\dagger
\Delta_i\, ,
\end{equation}
where $U_R$ is the $3 \times 6$ matrix introduced in eq.~(\ref{UU1}).
The mass $m$ is arbitrary,
since only the off-diagonal terms of eq.~(\ref{1loop}),
which are finite and $m$-independent,
have physical significance.
If we choose $m$ to be the right-handed-neutrinos mass scale,
we see that the large logarithms $\ln \left( m_{1,2,3}^2 / m^2 \right)$
are suppressed by small mixing angles in $U_R$.
The flavour-changing neutral interactions
are weak because they are cubic in the Yukawa coupling constants.
This guarantees that our theory is viable.
For instance,
there is a stringent bound of order $10^{-12}$
on the branching ratio of $\mu^- \to e^- e^+ e^-$ \cite{groom}.
With the result of eq.~(\ref{1loop})
we can estimate that branching ratio to be of order
$Y^8 / \left( 16 \pi^2 G_F m_H^2 \right)^2$,
where $Y$ is a typical Yukawa coupling
and $m_H$ is a typical neutral-scalar mass.
Taking $m_H \sim 100$ GeV and $Y \sim 10^{-3}$--$10^{-2}$
($Y$ should be of order of a charged-lepton mass
divided by the Fermi scale),
we obtain a branching ratio of $10^{-18}$ or smaller,
quite safe when compared to the experimental bound.

Charged-scalar exchange
also induces the radiative decay $\mu^- \to e^- \gamma$,
which has an experimental upper bound on its branching ratio
of order $10^{-11}$ \cite{groom}.
As before,
the transition amplitude for this decay
has both light and heavy neutrinos in the loop.
There is also a chirality flip and,
therefore,
a mass insertion.
The calculation of $\mu^- \to e^- \gamma$ in the present theory
is not similar to the one in the context of the Zee model \cite{petcov},
but rather to the ones in the context of supersymmetric models---see
for instance \cite{hisano}.
The branching  ratio of $\mu^- \to e^- \gamma$ is suppressed in our model not
only by the fine-structure constant $\alpha$
and a product of four Yukawa couplings, but also due to a GIM mechanism
\cite{GIM}. A crude estimate of an upper bound on this branching ratio
is given by
$\alpha Y^4 / \left(\pi G_F^2 m_R^4 \right)$, where $m_R$ denotes the scale of
the Majorana mass terms of the right-handed neutrinos. 
This one-loop estimate is far below the present
experimental bound. A detailed account of this decay
and of other flavour-changing decays
in the context of our framework will be presented in a forthcoming paper
\cite{future}.

\section{Maximal atmospheric mixing} \label{maxatm}

Within the theory described in the previous section
it is possible to implement maximal atmospheric mixing naturally.
A key to this possibility is the conservation of the lepton numbers $L_e$,
$L_\mu$,
and $L_\tau$ in the Yukawa interactions,
making the charged-lepton mass matrix diagonal;
we have to worry only about the form of $M_R$ and $M_D$.
We consider $n_H = 3$ and introduce two $Z_2$ symmetries:
\begin{eqnarray}
 & Z_2^{(1)}\!: &
\nu_{\mu R} \leftrightarrow \nu_{\tau R}\, , \;
D_\mu \leftrightarrow D_\tau\, , \;
\mu_R \leftrightarrow \tau_R\, , \;
\phi_3 \to -\phi_3\, ;
\label{Z2} \\
 & Z_2^{(2)}\!: &
\mu_R \to - \mu_R\, , \;
\tau_R \to - \tau_R\, , \;
\phi_2 \to - \phi_2\, , \;
\phi_3 \to -\phi_3\, .
\label{Z2'}
\end{eqnarray}
Fields not appearing in these equations transform trivially.
Let us motivate these choices.
Because of $Z_2^{(1)}$ we have
\begin{equation}
\left( M_R \right)_{e \mu} = \left( M_R \right)_{e \tau}
\quad {\rm and} \quad
\left( M_R \right)_{\mu \mu} = \left( M_R \right)_{\tau \tau}\, .
\end{equation}
As $\phi_2$ and $\phi_3$ change sign under $Z_2^{(2)}$,
only $\phi_1$ has Yukawa couplings to the neutrinos:
\begin{equation}\label{LYn}
\mathcal{L}_{\rm Y} (\nu_R) =
- \frac{\sqrt{2}}{v_1}
\left( \begin{array}{cc} \varphi_1^0, & - \varphi_1^+ \end{array} \right)
\left[ a \bar\nu_{eR} D_e +
b \left( \bar\nu_{\mu R} D_\mu + \bar\nu_{\tau R} D_\tau \right) \right]
+ \mbox{h.c.}\, ,
\end{equation}
cf.\ eq.~(\ref{lyukawa}).
Thus,
\begin{equation} \label{MD}
M_D = \mbox{diag} \left( a, b, b \right)
\end{equation}
has identical $\mu$ and $\tau$ entries,
once again because of $Z_2^{(1)}$.
As a consequence,
the light-neutrino Majorana mass matrix
has the same structure as $M_R$:
\begin{equation}\label{Mnu}
\mathcal{M}_\nu =
\left( \begin{array}{ccc}
x & y & y \\ y & z & w \\ y & w & z
\end{array} \right).
\end{equation}
Maximal atmospheric neutrino mixing and $U_{e3} = 0$
immediately follow from this structure of $\mathcal{M}_\nu$.\footnote{This
structure of $\mathcal{M}_\nu$ in the basis where the charged-lepton mass
matrix is diagonal has been suggested by several authors,
e.g.\ \cite{ma,BGS01,lam}.
We stress that in our case this structure results from a symmetry,
i.e.,
we have a {\em model\/} and not just a {\em texture\/} for $\mathcal{M}_\nu$.}
Using an adequate phase convention,
cf.\ eq.~(\ref{V}),
to transit from the unitary matrix $V$ which diagonalizes $\mathcal{M}_\nu$
to the neutrino mixing matrix $U$,
we obtain
\begin{equation}\label{U}
U = \left( \begin{array}{ccc}
\cos \theta & \sin \theta & 0 \\
\sin \theta/\sqrt{2} & -\cos \theta/\sqrt{2} & -1/\sqrt{2} \\
\sin \theta/\sqrt{2} & -\cos \theta/\sqrt{2} &  1/\sqrt{2}
\end{array} \right).
\end{equation}
Indeed,
as the charged-lepton mass matrix is diagonal within our general framework,
the matrix $U$ is already the neutrino mixing matrix.
The third column of $U$ in eq.~(\ref{U})
is an eigenvector of the mass matrix $\mathcal{M}_\nu$ of eq.~(\ref{Mnu}).
The form of the other two columns of $U$ follows
from the form of the third column.
At tree level the atmospheric mixing angle is exactly $45^\circ$,
whereas the solar mixing angle $\theta$ is free.
Without loss of generality we may assume $m_1 < m_2$
and $0^\circ < \theta < 90^\circ$.

Up to now a single Higgs doublet $\phi_1$ was needed.
Coming to the charged-lepton masses,
we obviously have to break $Z_2^{(1)}$ in order to avoid $m_\mu = m_\tau$.
This is achieved by introducing two more doublets $\phi_2$ and $\phi_3$,
one of them being even and the other one being odd under $Z_2^{(1)}$.
These doublets must not couple to $\nu_R$
because we want to avoid destruction of the form of $M_D$ in eq.~(\ref{MD});
this is the rationale for introducing the symmetry $Z_2^{(2)}$.
We obtain the following Yukawa couplings to the charged leptons:
\begin{eqnarray}
\mathcal{L}_{\rm Y} (\ell_R) &=&
- \frac{\sqrt{2} m_e}{v_1^\ast} \left( \begin{array}{cc}
\varphi_1^-, & {\varphi_1^0}^\ast \end{array} \right) \bar e_R D_e
\nonumber \\
& & - \sqrt{2} d \left( \begin{array}{cc}
\varphi_2^-, & {\varphi_2^0}^\ast \end{array} \right)
\left( \bar \mu_R D_\mu + \bar \tau_R D_\tau \right)
\nonumber \\
& & - \sqrt{2} d' \left( \begin{array}{cc}
\varphi_3^-, & {\varphi_3^0}^\ast \end{array} \right)
\left( \bar \mu_R D_\mu - \bar \tau_R D_\tau \right) +
\mbox{h.c.}\, ,
\label{LYl}
\end{eqnarray}
yielding
\begin{eqnarray}
m_\mu &=& \left| d v_2^\ast + d' v_3^\ast \right|, \\
m_\tau &=& \left| d v_2^\ast - d' v_3^\ast \right|.
\end{eqnarray}
This allows for $m_\mu \neq m_\tau$,
with some finetuning in order to obtain $m_\tau \gg m_\mu$.
Such a finetuning is,
anyway,
needed even within the SM.

We need to check that the Higgs potential is such
that it does not possess any U(1) symmetries
apart from the weak hypercharge U(1)$_Y$.
In particular,
we easily see that all terms of the form
$\left( \phi_i^\dagger \phi_j \right)^2$ are allowed
by both $Z_2^{(1)}$ and $Z_2^{(2)}$;
therefore,
there are no unwanted Goldstone bosons
and all the physical-scalar masses can be made sufficiently heavy,
of the order of the weak scale.
Notice that both $Z_2^{(1)}$ and $Z_2^{(2)}$ are broken spontaneously,
but at tree level this breaking is not felt in the neutrino sector.

In our model the seesaw mechanism is operative.
If we assume $a$ and $b$ to be of order $m^D_\nu \sim 0.1$--$1$ GeV
(where $m^D_\nu$ is a typical charged-lepton mass),
and if $m_R^2$ is the order of magnitude
of the eigenvalues of $M_R^* M_R$,
then we obtain the order-of-magnitude estimate
$\sqrt{\Delta m^2_\mathrm{atm}} \sim (m^D_\nu)^2/m_R$,
which gives $m_R \sim 10^8$--$10^{10}$ GeV.

Let us deal a bit longer with $\mathcal{M}_\nu$ of eq.~(\ref{Mnu}).
The neutrino masses are given by
\begin{equation}
m_3 = \left| z - w \right| \label{m3}
\end{equation}
and
\begin{equation}
m_{1, 2}^2 = \frac{1}{2} \left[
|x|^2 + 4|y|^2 + |z+w|^2 \mp \sqrt{
\left( |x|^2 + 4|y|^2 + |z+w|^2 \right)^2
- 4 \left| x (z+w) - 2 y^2 \right|^2}
\right].
\end{equation}
The solar mixing angle is expressed as
\begin{equation} \label{theta}
\tan 2\theta = 2\sqrt{2}\, \frac{|x^*y+y^*(z+w)|}{|z+w|^2-|x|^2}\, .
\end{equation}
Notice in particular that the only physical phase in $\mathcal{M}_\nu$ is
\begin{equation}
\beta = \arg x + \arg (z+w) - 2 \arg y\, .
\end{equation}
Indeed,
we have
\begin{equation} \label{theta1}
\tan 2\theta = 2\sqrt{2} \left| y \right|
\frac{\left| e^{i\beta} |z+w| + |x| \right| }{|z+w|^2-|x|^2}
\end{equation}
and
\begin{equation}\label{dsolar}
\Delta m^2_\odot
= \sqrt{\left( \left| z+w \right|^2 - \left| x \right|^2 \right)^2 +
8 \left| y \right|^2
\left| e^{i\beta} \left| z+w \right| + \left| x \right| \right|^2} \,.
\end{equation}
Combining these two equations leads to \cite{BGS01}
\begin{equation} \label{dsolar1}
\Delta m_\odot^2\, \cos 2\theta = |z+w|^2 - |x|^2\, ,
\end{equation}
where we have taken into account that,
by definition,
$\theta$ lies in the first quadrant.
This model does not allow to relate $\Delta m_\odot^2$
with $\Delta m_\mathrm{atm}^2$,
since $m_3$ is independent from $m_{1,2}$.

Given any values of $\Delta m_\odot^2$,
$\Delta m_\mathrm{atm}^2$,
and $\theta$,
they can be reproduced within the present model.
First we use eq.~(\ref{dsolar1}) to express $|z+w|$,
plug this expression into eq.~(\ref{theta}) and arrive at
\begin{equation} \label{sin2theta}
\sin 2\theta =  2\sqrt{2}\, \frac{|x||y|}{\Delta m^2_\odot}
\left| 1 + e^{i\beta} \sqrt{1 + \Delta m^2_\odot \cos 2\theta/|x|^2}
\right| \,.
\end{equation}
Obviously,
by choosing $|x|$ and $|y|$ we can achieve any desired value of $\theta$,
and by choosing $|z-w| = m_3$
we reproduce the experimental value of the atmospheric mass-squared difference.

Let us now address the question of whether one can still fit
any values of $\Delta m_\odot^2$,
$\Delta m_\mathrm{atm}^2$,
and $\theta$,
when $|x|$,
$|y|$,
$|z+w|$,
and $|z-w|$ are all assumed to be of order $\sqrt{\Delta m_\mathrm{atm}^2}$.
A glance at eq.~(\ref{sin2theta}) shows that this is impossible for
$|1 + e^{i\beta}| \gg \Delta m^2_\odot/\Delta m_\mathrm{atm}^2$.
However,
if we assume for simplicity that $e^{i\beta} = -1$,
then we have
\begin{equation}\label{tan}
\tan 2\theta \simeq \sqrt{2}\, \frac{|y|}{|x|}
\end{equation}
and,
in general,
we have a large solar neutrino mixing.
Naturally,
with our order-of-magnitude assumption on the absolute values,
finetuning is required to make the solar mass-squared difference small.
This finetuning is expressed,
e.g.,
by eq.~(\ref{dsolar1}) \cite{BGS01}.

At this stage it is appropriate to ask
whether there is any reason for having $e^{i\beta} = -1$.
Since we need a real phase factor,
we explore the effect of $CP$ invariance:
\begin{equation} \label{CP}
CP: \quad D_\alpha \to \gamma^0 C \bar D_\alpha^T\, , \;
\ell_R \to \gamma^0 C \bar \ell_R^{\,T} \, , \;
\nu_R \to \gamma^0 C \bar \nu_R^T \, , \;
\phi_j \to \phi_j^*\, .
\end{equation}
This $CP$ symmetry leads to real Yukawa-coupling matrices
$\Gamma_j$ and $\Delta_j$,
while $M_R$ is imaginary.
With eq.~(\ref{seesaw}) and the definition of $\beta$
it is easy to check that\footnote{Note that
we have the requirement
$\left( M_R \right)_{\mu \mu}^2 \neq \left( M_R \right)_{\mu \tau}^2$.
Indeed,
for $\left( M_R \right)_{\mu \mu} = \left( M_R \right)_{\mu \tau}$
the matrix $M_R$ is singular,
while for $\left( M_R \right)_{\mu \mu} = - \left( M_R \right)_{\mu \tau}$
we have $x=0$ and the phase $\beta$ is not defined.}
\begin{eqnarray}
e^{i\beta} &=& 
\mbox{sign} \left\{ \left( M_R \right)_{ee} \left[
\left( M_R \right)_{\mu \mu} + \left( M_R \right)_{\mu \tau}
\right] / \left( M_R \right)_{e \mu}^2 \right\}
\nonumber\\ &=&
- \mbox{sign} \left\{ \left( M_R \right)_{ee} \left[
\left( M_R \right)_{\mu \mu} + \left( M_R \right)_{\mu \tau}
\right] \right\}.
\end{eqnarray}
It is thus natural to have $\beta = \pi$ if we invoke $CP$ invariance.
In this case,
with all the matrix elements of $\mathcal{M}_\nu$
of order $\sqrt{\Delta m^2_\mathrm{atm}}$,
the only remaining finetuning which is required is to make
$|z+w|-|x|$ sufficiently small,
see eq.~(\ref{dsolar}).

\section{Maximal solar mixing}
\label{maxsolar}

In this section we dispense
with the $Z_2^{(1)}$ and $Z_2^{(2)}$ symmetries of the previous section.
Our present aim is to show that,
if one assumes the three individual lepton numbers $L_e$,
$L_\mu$,
and $L_\tau$ to be broken down at some high scale $\bar m_R$
to their linear combination $\bar L = L_e - L_\mu - L_\tau$,
and one further assumes that $\bar L$ only gets broken
at a much lower scale $m_R$ \cite{lavoura00},
then approximate maximal solar neutrino mixing follows.
The crucial point is that the assumption
of $\bar L$ conservation in between the two high scales $\bar m_R$ and $m_R$
is natural in the technical sense,
since $\bar L$ is a symmetry,
and therefore maximal solar mixing is a natural option
in the context of our framework.

Let us define the small dimensionless parameter
$\epsilon = m_R / \bar m_R \ll 1$.
The mass matrix $M_R$ then has the form
\begin{equation}\label{MRLbar}
M_R = \left( \begin{array}{ccc}
u & p / \epsilon & q / \epsilon \\
p / \epsilon & r & t \\
q / \epsilon & t & s
\end{array} \right),
\end{equation}
where $u$,
$p$,
$q$,
$r$,
$s$,
and $t$ are assumed to be all of order of magnitude $m_R$.
Indeed,
intermediate $\bar L$ conservation implies that the $(e, \mu)$
and $(e, \tau)$ entries of $M_R$ are the only ones
to be of order $\bar m_R$.
Taking into account that
$M_D = {\rm diag} \left( a, b, c \right)$ is diagonal,
we find that the seesaw neutrino mass matrix of eq.~(\ref{seesaw}) is
\begin{equation} \label{marvelous}
{\cal M}_\nu = \frac{1}
{p^2 s + q^2 r - 2 p q t + \epsilon^2 u \left( t^2 - r s \right)}
\left( \begin{array}{ccc}
\epsilon^2 a^2 \left( r s - t^2 \right) &
\epsilon a b \left( q t - p s \right) &
\epsilon a c \left( p t - q r \right) \\
\epsilon a b \left( q t - p s \right) &
b^2 \left( \epsilon^2 u s - q^2 \right) &
b c \left( p q - \epsilon^2 u t \right) \\
\epsilon a c \left( p t - q r \right) &
b c \left( p q - \epsilon^2 u t \right) &
c^2 \left( \epsilon^2 u r - p^2 \right)
\end{array} \right).
\end{equation}
Notice that only the $(\mu, \mu)$,
$(\tau, \tau)$,
and $(\mu, \tau) = (\tau, \mu)$ matrix elements of ${\cal M}_\nu$
are not $\epsilon$-suppressed;
still,
\begin{equation} \label{vanishes}
\left( {\cal M}_\nu \right)_{\mu\mu}
\left( {\cal M}_\nu \right)_{\tau\tau}
-
\left( {\cal M}_\nu \right)_{\mu\tau}
\left( {\cal M}_\nu \right)_{\tau\mu}
=
\frac{- \epsilon^2 b^2 c^2 u}
{p^2 s + q^2 r - 2 p q t + \epsilon^2 u \left( t^2 - r s \right)}
\end{equation}
is suppressed by two powers of $\epsilon$,
and this fact is crucial in the following.

We shall perform an expansion in the small parameter $\epsilon$
and compute the subleading order in $\epsilon$ of the relevant quantities.
This is consistent with our use of the seesaw formula
of eq.~(\ref{seesaw});
indeed,
if one wanted to compute sub-subleading orders in the $\epsilon$ expansion
one would first need a better version  of the seesaw expansion
of ${\cal M}_\nu$ \cite{seesaw2}.

For the sake of simplicity we shall assume the matrix ${\cal M}_\nu$
of eq.~(\ref{marvelous}) to be real.\footnote{\label{CPfoot}By imposing
the $CP$ symmetry of eq.~(\ref{CP}),
and with the help of a phase transformation,
the imaginary matrix $M_R$ can be made real while keeping $M_D$ real.
Then ${\cal M}_\nu$ will be real too.}
This allows us to diagonalize it
by simply looking for its eigenvalues and eigenvectors.
Thus,
\begin{equation}
U^T {\cal M}_\nu U = {\rm diag}
\left( \eta_1 m_1, \eta_2 m_2, \eta_3 m_3 \right),
\label{jsker}
\end{equation}
where the $\eta_k$ may be either $+1$ or $-1$,
and $U$ is the neutrino mixing matrix as usual.
If ${\cal M}_\nu$ was not real
one would first have to diagonalize ${\cal M}_\nu {\cal M}_\nu^\ast$,
and that would of course be quite more tedious.
One of the eigenvalues (say, $\lambda_0$)
of ${\cal M}_\nu$ is of order $(m^D_\nu)^2 / m_R$,
and the other two (we call them $\lambda_\pm$) are smaller,
of order $(m^D_\nu)^2 / \bar m_R$,
because of eq.~(\ref{vanishes}).
The eigenvalues $\lambda_\pm$
fulfill $|\lambda_+| = |\lambda_-|$
to leading order in $\epsilon$,
which means that $\Delta m^2_\odot$ vanishes to that order.
Explicitly,
to subleading order in $\epsilon$
\begin{eqnarray}
\lambda_0 &=& \frac{- F^2}{p^2 s + q^2 r - 2 p q t}\, , \\
\lambda_\pm &=& \pm \frac{\epsilon a b c}{F}
+ \frac{\epsilon^2 \left( a^2 b^4 q^2 s + a^2 c^4 p^2 r + 2 a^2 b^2 c^2 p q t
+ b^2 c^2 F^2 u \right)}{2 F^4}\, ,
\end{eqnarray}
where
\begin{equation}
F \equiv \sqrt{b^2 q^2 + c^2 p^2}\, .
\end{equation}
It follows that
\begin{eqnarray}
\frac{\Delta m^2_\odot}{\Delta m^2_{\rm atm}} &\simeq&
\frac{\left| \lambda_+^2 - \lambda_-^2 \right|}{\lambda_0^2}
\nonumber \\ &\simeq&
\left| \frac{2 \epsilon^3 a b c \left( p^2 s + q^2 r - 2 p q t \right)^2
\left( a^2 b^4 q^2 s + a^2 c^4 p^2 r + 2 a^2 b^2 c^2 p q t
+ b^2 c^2 F^2 u \right)}{F^9} \right|
\nonumber \\ & &
\end{eqnarray}
is of order $\epsilon^3$.
Maximal solar neutrino mixing is allowed at 90\% CL in the LOW solution
with $\Delta m^2_\odot \sim 10^{-7}$ eV$^2$,
giving the estimate $\epsilon^3 \sim 10^{-4}$.
In this case $\epsilon$ is small enough for a sensible expansion.
Approximate maximal mixing is also possible in the LMA MSW solution,
but then only at 99\% CL \cite{GG};
in that case $\epsilon^3 \sim 10^{-2}$ is rather large
and the two scales $m_R$ and $\bar m_R$ are not so clearly separated.

To subleading order in $\epsilon$,
the normalized eigenvector corresponding to the eigenvalue $\lambda_0$ is
\begin{equation}
\left( \begin{array}{c}
\epsilon a \left[ \left( b^2 q^2 - c^2 p^2 \right) t
+ p q \left( c^2 r - b^2 s \right) \right] / F^3 \\
- b q / F \\ c p / F
\end{array} \right),
\label{udoep}
\end{equation}
and the ones corresponding to the eigenvalues $\lambda_\pm$ are
\begin{equation}
\left( \begin{array}{c}
{\displaystyle \frac{1}{\sqrt{2}} \left[ 1 \pm \epsilon \left(
\frac{a b c p q t}{2 F^3} + \frac{a c^3 p^2 r}{4 b F^3}
+ \frac{a b^3 q^2 s}{4 c F^3} - \frac{b c u}{4 a F} \right) \right]
} \\*[3mm]
{\displaystyle
\mp \frac{c p}{\sqrt{2} F} \left[ 1 \pm {\rm O} \left(
\epsilon \right) \right]
} \\*[3mm]
{\displaystyle
\mp \frac{ b q}{\sqrt{2} F} \left[ 1 \pm {\rm O} \left(
\epsilon \right) \right]
}
\end{array} \right),
\end{equation}
respectively.
It follows that,
to leading order,
the mixing matrix $U$ is
\begin{equation}
U = \left( \begin{array}{ccc}
1 / \sqrt{2} & 1 / \sqrt{2} & 0 \\
\sin{\psi} / \sqrt{2} & - \sin{\psi} / \sqrt{2}
& - \cos{\psi} \\
\cos{\psi} / \sqrt{2} & - \cos{\psi} / \sqrt{2}
& \sin{\psi}
\end{array} \right),
\quad {\rm with} \quad \sin{\psi} \equiv c p /F\, , \
\cos{\psi} \equiv b q / F\, .
\end{equation}
Thus,
$U_{e 3} = 0$ and maximal solar mixing hold at leading order.
At subleading order
\begin{eqnarray}
U_{e3} &=& \epsilon\, \frac{a \left[ \left( b^2 q^2 - c^2 p^2 \right) t
+ p q \left( c^2 r - b^2 s \right) \right]}{F^3}\, , \\
\sin^2{2 \theta} &=& 1 - \epsilon^2 \left(
\frac{a b c p q t}{F^3} + \frac{a c^3 p^2 r}{2 b F^3}
+ \frac{a b^3 q^2 s}{2 c F^3} - \frac{b c u}{2 a F} \right)^2.
\end{eqnarray}
It is worth emphasizing that $\Delta m^2_\odot / \Delta m^2_{\rm atm}$
is of order $\epsilon^3$
while $1 - \sin^2{2 \theta}$ is proportional to $\epsilon^2$ only.
This means that,
for solutions of the solar neutrino problem
with a higher value of $\Delta m^2_\odot / \Delta m^2_{\rm atm}$,
such as the LMA MSW solution,
we may allow $\sin^2{2 \theta}$ to be further away from unity.
This gives us extra room in the fitting of the experimental data
within the context of our model.

We also remark that some models based on $\bar L$ symmetry \cite{lavoura00}
predict a massless neutrino ($m_3 = 0$)
which has zero component along the $\nu_e$ direction ($U_{e3} = 0$).
The present model,
on the contrary,
allows for $U_{e3}$ to differ from 0 substantially
(as it is only of order $\epsilon$)
and,
moreover,
it predicts $m_3$ to be much larger than $m_1$ and $m_2$,
instead of $m_3 = 0$.
Note that our model displays $m_2 - m_1 \ll (m_1 + m_2)/2$, 
contrary to the orthodox hierarchical pattern
of masses, where $m_1 \ll m_2$.

\section{Bimaximal mixing} \label{bimax}

We may combine the assumptions
of sections~\ref{maxatm} and \ref{maxsolar}
to obtain a scheme with {\em natural\/} bimaximal mixing.
In that scheme $M_D = {\rm diag} \left( a, b, b \right)$
as in eq.~(\ref{MD}) and
\begin{equation}
M_R = \left( \begin{array}{ccc}
u & p / \epsilon & p / \epsilon \\
p / \epsilon & r & t \\
p / \epsilon & t & r
\end{array} \right),
\end{equation}
with $\epsilon$ much smaller than 1,
in analogy to eq.~(\ref{MRLbar}).
Then,
assuming once again $M_D$ and $M_R$ to be real
for the sake of simplification (see footnote~\ref{CPfoot}), we obtain
\begin{eqnarray}
\lambda_0 &=& \frac{b^2}{t-r}\, , \\
\lambda_\pm &=& \pm \frac{\epsilon a b}{\sqrt{2} p}
+ \frac{\epsilon^2 \left[ a^2 \left( r+t \right) + b^2 u \right]}{4 p^2}\, ,
\end{eqnarray}
and
\begin{equation}
\sin^2{2 \theta} = 1 - \frac{\epsilon^2}{8 p^2}
\left[ \frac{a}{b} \left( r+t \right) - \frac{b}{a}\, u \right]^2.
\end{equation}

Maximal atmospheric neutrino mixing
and $U_{e3} = 0$ are exact results in this case.
Indeed,
the full $6 \times 6$ neutrino mass matrix
\begin{equation}
\mathcal{M}_{D+M} = \left( \begin{array}{cccccc}
0 & 0 & 0 & a & 0 & 0 \\
0 & 0 & 0 & 0 & b & 0 \\
0 & 0 & 0 & 0 & 0 & b \\
a & 0 & 0 & u & p / \epsilon & p / \epsilon \\
0 & b & 0 & p / \epsilon & r & t \\
0 & 0 & b & p / \epsilon & t & r \\
\end{array} \right)
\end{equation}
has an eigenvalue
\begin{equation}\label{eigen}
\left( t - r \right)
\left( \sqrt{1 + \sigma^2} - 1 \right) / 2
\simeq \frac{b^2}{t - r}\, ,
\end{equation}
where
\begin{equation}
\sigma \equiv \frac{2 b}{r - t}\, .
\end{equation}
The above eigenvalue is seesaw-suppressed
and its absolute value corresponds to $m_3$.
To the exact eigenvalue (\ref{eigen}) corresponds the exact eigenvector of
$\mathcal{M}_{D+M}$
\begin{equation}
\frac{1}{2 \left( 1 + \sigma^2 \right)^{1/4}}
\left( \begin{array}{c}
0 \\ \sigma / \eta \\ - \sigma / \eta \\ 0 \\ - \eta \\ \eta
\end{array} \right),
\quad {\rm with}\ \eta \equiv \sqrt{\sqrt{1 + \sigma^2} - 1}
\simeq \frac{\sigma}{\sqrt{2}}\, .
\end{equation}
Note that both the eigenvalue and the eigenvector
depend only on $b$ and on $r - t$;
in particular,
they depend neither on $a$ nor on $p$.

\section{Conclusions} \label{conclusions}

In this paper we have considered the extension of the SM
with three right-handed neutrino singlets
with large Majorana mass terms responsible for the seesaw mechanism.
Upon this standard scenario
we have imposed the separate conservation
of the family lepton numbers $L_e$,
$L_\mu$,
and $L_\tau$ in the Yukawa couplings,
such that these lepton numbers are softly broken
solely by the Majorana mass terms.
In this way,
at tree level the charged-lepton mass matrix is automatically diagonal
and lepton mixing originates exclusively
in the Majorana mass matrix $M_R$.

This makes it relatively easy to impose a discrete symmetry
which enforces $U_{e3} = 0$ and maximal atmospheric neutrino mixing;
for this we need three Higgs doublets.
An alternative model,
with maximal solar mixing,
is obtained when there are two different scales in $M_R$ such that,
at the higher scale,
the individual lepton numbers $L_e$,
$L_\mu$,
and $L_\tau$ are broken down to $\bar L =L_e - L_\mu - L_\tau$;
whereas at the lower scale $\bar L$ is also softly broken.
In this case one Higgs doublet suffices.
The two models can easily be combined
if one wants to obtain bimaximal mixing.

In the model with maximal atmospheric mixing
the ratio $\Delta m^2_\odot/\Delta m^2_\mathrm{atm}$
has to be fitted by means of a finetuning.
In the model with maximal solar mixing
there is a relationship between the deviation of $\theta$ from $45^\circ$
and the ratio of the mass-squared differences,
namely $1-\sin^2 2\theta \sim \epsilon^2$ and
$\Delta m^2_\odot/\Delta m^2_\mathrm{atm} \sim \epsilon^3$,
where $\epsilon$ is the ratio of the two scales in $M_R$.

The mechanisms for maximal neutrino mixing discussed in this paper
are extremely simple
and require only a minimal extension of the SM
with right-handed singlets $\nu_R$.
Apart from the possibility of obtaining maximal neutrino mixing
in a natural way,
the violation of $L_e$,
$L_\mu$,
and $L_\tau$ exclusively by the large Majorana masses of the $\nu_R$
constitutes in itself an interesting scenario,
where deviations from the conservation of the family lepton numbers
are calculable,
finite,
and controlled by $M_R$ only.

\end{document}